\documentclass[aps,reprint,showpacs,amsmath,amssymb,pra,longbibliography,superscriptaddress]{revtex4-2}
\usepackage{amsmath,amssymb,latexsym,epsfig,graphics,epsf,siunitx}

\begin{document}
\title{On the Role of Internal Degrees of Freedom in Structural Relaxation of Ring--Tail Structured Liquids Across Temperature Regimes}

\author{Rolf Zeißler}
\affiliation{Institute for Condensed Matter Physics, Technical University of Darmstadt, D-
64289 Darmstadt, Germany}
\author{Sandra Krüger}
\affiliation{Institute for Condensed Matter Physics, Technical University of Darmstadt, D-
64289 Darmstadt, Germany}
\author{Robin Horstmann}
\affiliation{Institute for Condensed Matter Physics, Technical University of Darmstadt, D-
64289 Darmstadt, Germany}
\author{Till Böhmer}
\affiliation{Institute of Frontier Materials on Earth and in Space, German Aerospace Center,
D-51147 Cologne, Germany}
\author{Michael Vogel}
\affiliation{Institute for Condensed Matter Physics, Technical University of Darmstadt, D-
64289 Darmstadt, Germany}
\author{Thomas Blochowicz}
\affiliation{Institute for Condensed Matter Physics, Technical University of Darmstadt, D-
64289 Darmstadt, Germany}

\begin{abstract}
We investigate how anisotropic molecular rotation and internal molecular flexibility influence liquid dynamics in 1-phenylalkanes. To this end, we combine depolarized dynamic light scattering, nuclear magnetic resonance spectroscopy and molecular dynamics simulations. Our results show that anisotropic rotations and internal molecular flexibility substantially contribute to structural relaxation in the liquid state. However, their influence diminishes on entering the supercooled-liquid regime, where the relaxation behavior develops towards the previously identified generic relaxation shape, likely due to the increasing cooperativity of rotational dynamics. Because 1-phenylalkanes are simple model systems with similarities to many other molecular liquids, this study suggests that effects of anisotropic rotation and internal flexibility are relevant in various liquids with similar molecular complexity, and provides a proof of concept for how these effects can be identified.
\end{abstract}

\maketitle

\section{Introduction}

Over the past few decades, numerous studies have investigated the spectral shape of relaxation processes in a wide range of glass-forming liquids using various experimental techniques.\cite{angell2000relaxation,lunkenheimer2000glassy,bohmer2001dynamics,bohmer2025spectral} Commonly, a clear distinction is made between small, rigid molecules and larger, flexible macromolecules such as polymers. For small molecules, the relaxation spectra, apart from possible contributions from secondary relaxations, are typically interpreted in terms of a single structural- or $\alpha$-relaxation peak, which corresponds to the cooperative molecular reorientation dynamics. Its dispersion is usually thought to reflect dynamic heterogeneity throughout the system\cite{lunkenheimer2000glassy} and the reorientation most often is considered to be isotropic in good approximation\cite{Hinze1998a, Boehmer1998a}.

In polymers, by contrast, the structural relaxation peak corresponds to local segmental rotations. Additional contributions involve the motion of multiple segments up to the entire polymer chain, as, e.g., seen in the dielectric normal modes \cite{schonhals2003molecular}. Moreover, the local segmental relaxation can be influenced by these large-scale motions through mechanisms such as chain connectivity, resulting in a wide variety of structural relaxation peak shapes.\cite{paluch2010sub, abou2010rouse, hintermeyer2008molecular}

Besides flexible macromolecules, semi-rigid macromolecules have also become the focus of recent studies. Rams-Baron et al.\cite{rams2020broadband, rams2021complex, blazytko2024influence, raj2024dual, rams2024insight} have demonstrated in a series of studies on so-called sizable glass formers, which are molecules composed of a small polar group attached to a large, rigid, and non-polar core, that the position of the polar group, i.e., the orientation of the dipole moment relative to the molecular axes, significantly influences the frequency dispersion of the structural relaxation peak. This effect arises primarily from the anisotropic rotation of the asymmetrically shaped molecules.

Naturally, there must be a continuous transition from liquids composed of small rigid molecules to those comprising various types of macromolecules. This transition occurs through intermediate-sized molecules with varying degrees of flexibility. It is the goal of the present contribution to ascertain to what extent the intramolecular flexibility and possible anisotropic shape of such molecules affects the shape of their relaxation spectra.

It is important to note that, even for relatively small and rigid molecular glass formers, the spectral shape of the structural relaxation is still subject to controversial debate. The traditional view is that at high temperatures, all molecules undergo rotational diffusion characterized by a single, well-defined relaxation time. This results in an approximately mono-exponential rotational correlation function or, in frequency-domain measurements, a symmetric Debye peak. As the temperature is lowered below the melting point, however, the dynamics is believed to become heterogeneous, i.e., different regions within the liquid exhibit distinct characteristic timescales of molecular motion. This change in the distribution of relaxation times is thought to lead to a progressive broadening of the relaxation peak as the system approaches the glass transition.\cite{ediger1996supercooled, bohmer1998nature, sillescu1999heterogeneity, ediger2000spatially, richert2002heterogeneous,cavagna2009supercooled}

However, in recent years, it has become clear that this simple picture often does not reflect the experimental observations. On the one hand, deviations from the ideal Debye shape are frequently observed even at temperatures well above the melting point. More specifically, an asymmetric broadening of the relaxation peak is found, which is individual for different liquids.\cite{lunkenheimer2008broadband, lunkenheimer2000glassy, schneider2000scaling, schmidtke2014relaxation, schmidtke2013reorientational, rossler2024glass, rossler2025relaxation, bohmer2025spectral} On the other hand, recent studies have shown that, when using experimental methods that are insensitive to cross correlations between different molecules, the structural relaxation peak in the supercooled regime near the glass transition temperature exhibits a generic shape across various types of liquids.\cite{pabst2021generic, bohmer2025spectral} This again raises the question regarding the molecular origin of the spectral shape of structural relaxation and its evolution with temperature.

The influence of anisotropic rotation and internal molecular flexibility on the reorientation dynamics has intensively been investigated by nuclear magnetic resonance (NMR) spectroscopy. \cite{suchanski1999nuclearmagnetic, suchanski1998nmr, szczesniak1997dynamics, min199913c,qi2000slow, kim1995high, becher2021nuclear, hinze1995anisotropic} However, it has rarely been studied how these factors affect structural relaxation as probed by non-site-specific techniques, e.g., depolarized dynamic light scattering (DDLS) or broadband dielectric spectroscopy (BDS). Notable exceptions are studies on alcohols where hydrogen bonding can support dynamic separation of different molecular moieties, leading to a broadening of the structural relaxation peak as measured by BDS or DDLS \cite{bohmer2022glassy,bohmer2023revealing,cheng2024hydrogen,arrese2017non}. 

Here, we explore the influence of anisotropic rotation and intramolecular flexibility on the structural relaxation of liquid 1-phenylalkanes by combining DDLS, NMR, and molecular dynamics (MD) simulations. Our results show that, even for relatively small molecules, anisotropic reorientation dynamics and internal degrees of freedom have a significant impact on the spectral shape of structural relaxation in the liquid state, while their impact diminishes in the supercooled regime. We argue that this crossover occurs due to the increasing cooperativity of rotational dynamics upon cooling. Due to the simple molecular structure of liquid 1-phenylalkanes, which share common features with many other molecular liquids, our findings may help to shed light on the above mentioned open questions concerning the spectral shape of structural relaxation and specifically, on the transition from diverse relaxation shapes encountered in the liquid regime above the melting point to the generic structural relaxation observed near the glass transition.

\section{Experimental Details}
\subsection{Samples}
We investigate five 1-phenylalkanes with varying length of the alkyl chain. In the following they will be referred to in terms of their alkyl chain length $n$. 1-phenylbutane ($n=4$) and 1-phenylhexane ($n=6$) were purchased from Thermo Scientific, 1-phenyloctane ($n=8$) was purchased from Acros Organics and 1-phenyltridecane ($n=13$) and 1-phenylnonadecane ($n=19$) from TCI America. The chemical purities of the liquids were specified as $>98\,$\SI{}{\percent} by the suppliers. For $n=4$, $n=6$, and $n=8$, 1-phenylalkanes with selective $^{2}$H labeling of the phenyl ring were obtained from CDN Isotopes at chemical purities of $>99\,$\SI{}{\percent} and $>99\,$\SI{}{\percent} isotopic enrichment.

\subsection{Depolarized dynamic light scattering}
DDLS spectra were obtained using photon correlation spectroscopy (PCS) in the mHz to MHz frequency range, and with a multipass tandem Fabry–Perot interferometer (TFPI) from JRS Scientific Instruments in the MHz to THz range.

In the TFPI experiment the sample is contained in a cuvette, which is placed in an oven for measurements above room temperature or an optical cryostat for measurements below room temperature. For the oven, temperature accuracy is approximately 2\,K and for the cryostat 0.5\,K. The sample is irradiated by the beam of a Coherent Verdi V2 laser with a wavelength of 532\,nm and the spectral density $I(\nu)$ of the depolarized component of the light scattered from the sample is measured by the TFPI in backscattering geometry. The depolarized component of the scattered light is accessed by using two Glan-Thompson polarizers, one positioned in the beam path of the laser before it irradiates the sample, transmitting vertically polarized light, and one positioned in the path of the scattered light, transmitting horizontally polarized light (VH geometry). The imaginary part of the DDLS susceptibility $\chi^{\prime\prime}(\nu)$ is then calculated via the fluctuation dissipation theorem:
\begin{equation}
\chi^{\prime\prime}(\nu)=\frac{I(\nu)}{n(\nu,T)+1}
\label{eq:fdt}
\end{equation}
where $n(\nu,T)=(\mathrm{exp}(h\nu/k_{\mathrm{B}}T)-1)^{-1}$ is the Bose temperature factor.

In the PCS experiment the sample is contained in a home-built light scattering sample cell consisting of a piece of quartz glass tube with an outer diameter of 20\,mm, enclosed at both ends with aluminium lids, sealed by O-rings. This sample cell is then screwed on to the end of a coldfinger positioned in a vacuum chamber. Temperature accuracy of this setup is $\leq0.5$\,K. The sample is irradiated by the beam of a Cobolt Samba laser with a wavelength of 532\,nm and the fluctuating scattered intensity is measured over time by two Count T-100 avalanche photo diodes by Laser Components. From the fluctuating intensity, an intensity autocorrelation function $g_2(t)$ is obtained using a hardware correlator (ALV 7000). From the intensity autocorrelation function, the autocorrelation function of the electric field $g_1(t)$ is calculated via the Siegert relation $g_1(t)=\sqrt{(g_2(t)-1)/\Lambda}$, where the coherence factor $\Lambda$ has been determined to be 0.98. The PCS experiments where performed at a scattering angle of 90\textdegree\ in depolarized VH geometry, which was realized by two Glan-Thompson polarizers as in the TFPI experiment. $\chi^{\prime\prime}(\nu)$ was then obtained from Fourier-Laplace transformation of $g_1(t)$, employing the Filon algorithm.

\subsection{Nuclear magnetic resonance}
\label{sec:NMR}

$^2$H NMR spin-lattice relaxation (SLR) experiments were performed on 1-phenylalkanes with $^2$H labeling at the phenyl ring. They probe fluctuations of the orientation-dependent $^2$H quadrupolar frequencies:\cite{SchmidtRohrSpiess2012}
\begin{equation}
    \omega_\text{Q}=\pm \frac{\delta}{2}(3\cos^2(\theta) -1) \propto P_2(\cos \theta) \,.
\end{equation}
Here, $\theta$ is the angle between an external magnetic field $B_0$ and the C--D bond axis and the anisotropy parameter $\delta$ describes the strength of the quadrupolar interaction, which amounts to 2$\pi\cdot$ 135\,kHz for C--D bonds at phenyl rings. \cite{Rössler_CPL_1984} Thus, in our case of ring-labeled 1-phenylalkanes, $^{2}$H~NMR is specifically sensitive to the rotational dynamics of the phenyl ring.

The $^2$H SLR times $T_1$ are connected to the spectral density $J_2$ of the phenyl ring reorientation via\cite{Pound_PR_48}
\begin{align}\label{eq:bpp}
    \frac{1}{T_1} = \frac{2}{15}\delta^2 \left[J_2(\omega_\text{L})+4J_2(2\omega_\text{L})\right] \, .
\end{align}
where $\omega_\text{L}$ denotes the Larmor frequency. To account for a distribution of correlation times, we assume that the spectral density $J_2(\omega)$ has a Cole-Davidson (CD) shape:
\begin{equation}
    J_{\text{CD}}(\omega)=\frac{\sin\left[\beta_\text{CD}\arctan(\omega\tau_\text{CD})\right]}{\omega\left[1+(\omega\tau_\text{CD})^2\right]^{\beta_\text{CD}/2}} \,.
\end{equation}
The CD width parameter $\beta_\text{CD}$ is obtained from the height of the $T_1$ minimum and is used to determine correlation times $\tau_\text{CD}(T)$ from $T_1(T)$.\cite{bohmer2001dynamics}
From the obtained $\beta_\text{CD}$ and $\tau_\text{CD}$ values, peak correlation times were calculated for a direct comparison with DDLS and MD values.\cite{Beckmann_PR_88}
In the limit of fast dynamics ($\omega_\text{L}\tau\ll 1$), mean correlation times $\langle\tau\rangle$ can be calculated directly from $T_1$ without prior knowledge of the exact shape of $J_2(\omega)$ using
\begin{equation} \label{eq:HT-Naeherung}
    \frac{1}{T_1}=\frac{2}{3}\delta^2 \langle\tau\rangle \ .
\end{equation}

The $^2$H SLR measurements were performed using home-built spectrometers operating at $^2$H Larmor frequencies $\omega_\text{L}$ of 46.1 and 46.7\,MHz and applying the saturation-recovery or inversion-recovery methods with a solid-echo readout, where the echo delay was 20\,$\mu$s and the 90° pulse length amounted to about 2\,$\mu$s.

\subsection{Molecular dynamics simulations}
\label{subsec:MD}

\begin{figure}[h]
  \includegraphics[width=0.5\textwidth]{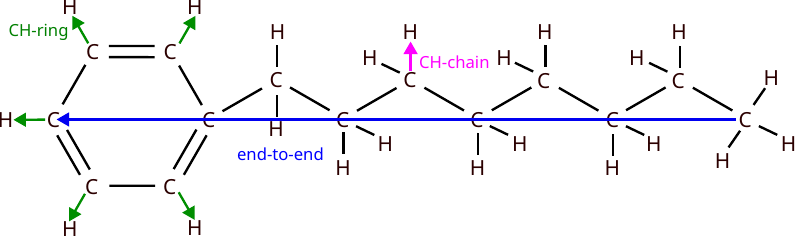}
  \caption{Structural formula of 1-phenyloctane ($n=8$). In the present MD simulations, we calculate rotational correlation functions for the end-to-end vector (blue), the C--H bond vectors of the alkyl chain (CH-chain, pink), and the C--H bond vectors of the phenyl ring (CH-ring, green).}
  \label{fig:figure1}
\end{figure}

In MD simulations, it is possible to calculate rotational correlation functions for any molecular moiety. For the present comparison with DDLS and NMR results for 1-phenylalkanes, it is particularly useful to consider the intramolecular vectors depicted in Fig.\ \ref{fig:figure1}. Specifically, we analyze the reorientation of the molecular end-to-end vector and of the C--H bonds in the alkyl chain (CH-chain) and in the phenyl ring (CH-ring), respectively. The end-to-end vector is predominantly sensitive to the rotation of the entire molecule, while discriminating between different C--H bonds allows us to selectively analyze the alkyl-tail and phenyl-ring dynamics. Moreover, for our analysis, it is advantageous to compute the rotational correlation functions $F_{2,k}(t)$ of the second Legendre polynomial $P_2$, explicitly, 
\begin{equation}
    F_{2,k}(t) \propto \langle P_2(\vec v_{k}(0)\cdot\vec v_{k}(t))\rangle
\end{equation}
where $\vec v_{k}$ is a normalized intramolecular vector and the pointed brackets $\langle\dots\rangle$ indicate averages over all molecules and various time origins. For straightforward comparison with DDLS and NMR results, we also obtain the corresponding dynamical susceptibilities from Fourier-Laplace transformation of $F_{2,k}(t)$. Finally, to analyze the orientational correlation between two different intramolecular vectors in the same molecule, $\vec v_{k}$ and $\vec v_{l}$, we use
\begin{equation}\label{eq:introt}
    C_{k,l}(t)=\langle P_2(\vec v_{k}(0)\cdot\vec v_{l}(0)) \cdot P_2(\vec v_{k}(t)\cdot\vec v_{l}(t))\rangle \,,
\end{equation}
which allows us to gain information on changes in the relative orientation of different intramolecular vectors through internal rotations.

The MD simulations were performed using the GROMACS package version 2023.3.\cite{GROMACS1,GROMACS2,GROMACS3,GROMACS4,GROMACS5} The parameterizations for all 1-phenylalkanes were obtained employing the small molecule topology generator STaGE\cite{STAGE, STAGE3}. The second generation General Amber Force Field GAFF2\cite{GAFF,ambertools23} was chosen with partial charges calculated using AM1-BCC.\cite{am1bcc1,am1bcc2} In addition to these flexible molecules, stiffened molecules were considered to study the role of the intramolecular flexibility. For stiffening, the dihedral potentials involving four carbon atoms, of which at least two belong to the alkyl chain, were multiplied by a factor of ten. 

In all MD simulations, the time step was $\Delta t=1$\,fs and periodic boundary conditions were applied with a cubic simulation box containing 1000 molecules. All hydrogen bonds were constrained utilizing LINCS.\cite{LINCS} Temperature and pressure were controlled employing the velocity-rescaling thermostat\cite{VR} and the stochastic cell rescaling barostat with a compressibility of $10^{-5}$\,bar$^{-1}$,\cite{CR} respectively. Non-bonded interactions were calculated up to a distance of 1\,nm, while long-range interactions were treated with dispersion corrections for Lennard-Jones interactions and with the particle-mesh Ewald (PME) method \cite{PME} for Coulomb interactions, using a Fourier spacing of 0.12\,nm.

Prior to the production runs in the NVT ensemble, we performed equilibration runs in the NPT ensemble to adjust the density at each studied temperature. In all simulations of the flexible molecules, the minimum length of the equilibration runs was 100 times the end-to-end vector rotational correlation time. 
For the simulations of the stiffened molecules, the starting configuration was taken from an equilibrated simulation of the original parametrization.

\section{Results and discussion}
\begin{figure}[ht!]
  \includegraphics[width=0.5\textwidth]{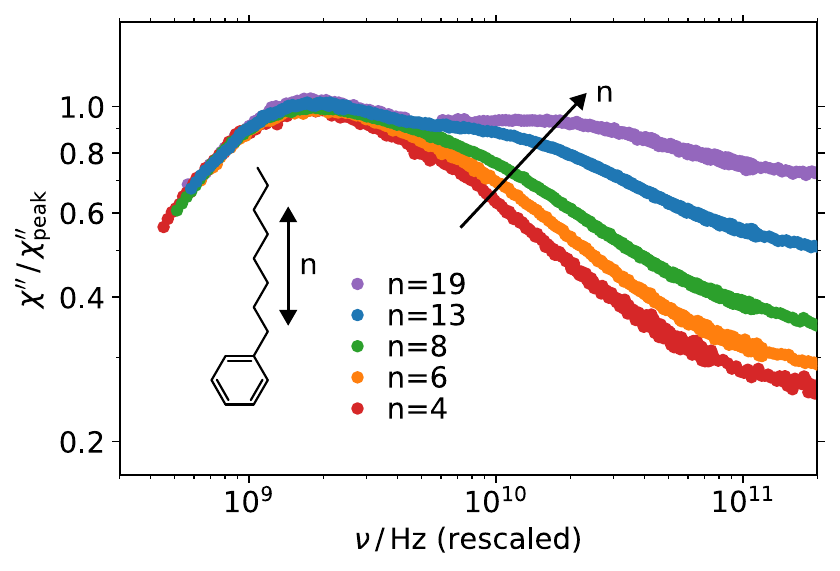}
  \caption{DDLS spectra of 1-phenylalkanes of varying length of the alkyl chain $n$ normalized to their maximum amplitude and mildly shifted in frequency for a detailed comparison of the peak shape. The temperatures of the measurements were chosen such that the peak frequencies amounted to $\sim$1.5\,GHz for all studied samples (see text for details).}
  \label{fig:figure2}
\end{figure}

Fig.\,\ref{fig:figure2} shows DDLS spectra of liquid 1-phenylalkanes with varying alkyl chain length $n$, normalized to their maximum amplitude and mildly shifted in frequency so that the low-frequency sides of the main relaxation peaks overlap. The spectra were obtained at 420\,K for $n=19$, 360\,K for $n=13$, 300\,K for $n=8$, 270\,K for $n=6$, and 240\,K for $n=4$. These temperatures were chosen such that the (unshifted) spectra exhibit similar peak frequencies of $\sim$1.5\,GHz, corresponding to peak relaxation times in the order of 100\,ps. In previous DDLS studies,\cite{zeissler2023influence, zeissler2025relation} these 1-phenylalkanes were investigated in moderate temperature ranges above their melting points.

As shown in Fig.\,\ref{fig:figure2}, the relaxation spectra exhibit bimodal relaxation peaks, with the degree of bimodality increasing with alkyl chain length $n$. In the aforementioned works,\cite{zeissler2023influence, zeissler2025relation} the fast component of the relaxation has been associated with phenyl ring rotation, and the slower component with the reorientation of the entire molecule. However, the extent to which this dynamic separation of the two relaxation contributions arises from internal rotation of the phenyl ring or rigid anisotropic rotation of the entire molecule could thus far not be determined experimentally and will be answered in the following by combination of DDLS, NMR and MD simulations.

\begin{figure}[ht!]
  \includegraphics[width=0.5\textwidth]{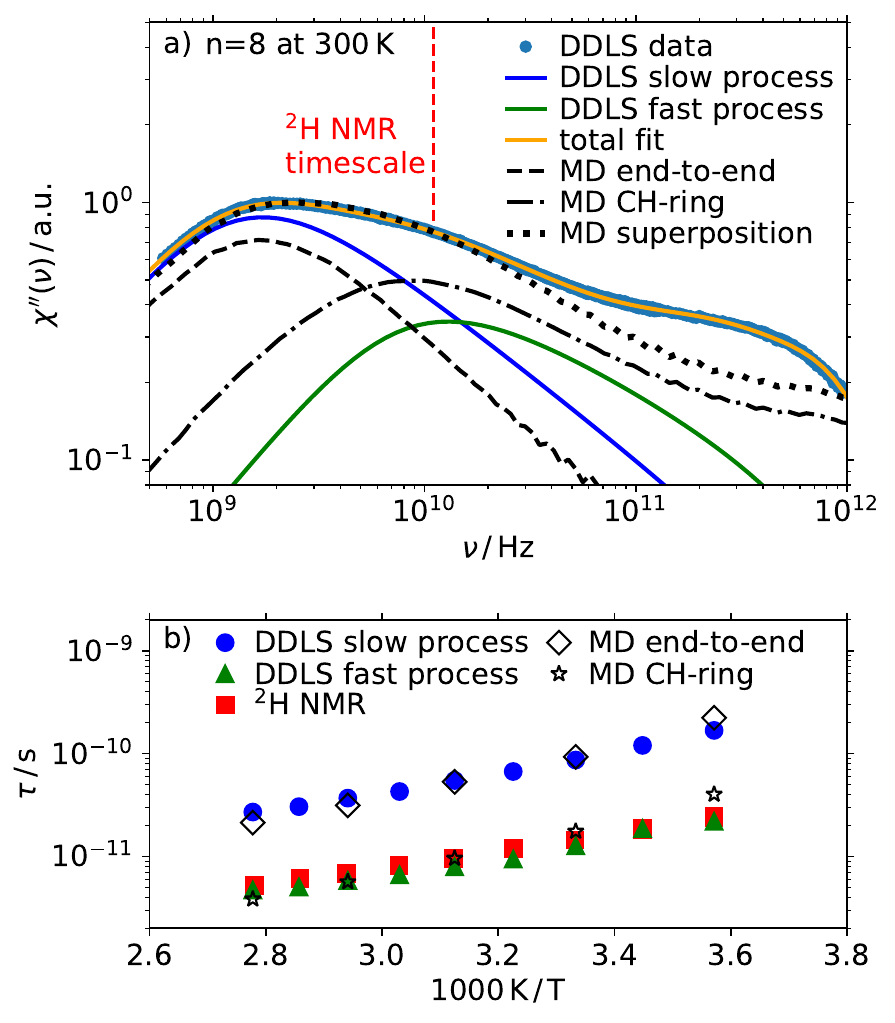}
  \caption{a) DDLS spectrum of 1-phenyloctane ($n=8$) at 300\,K. The solid orange curve is the total fit function developed in a previous work\cite{zeissler2023influence} and described in the text, while the solid blue and green curves are the two involved contributions. The vertical dashed line marks the frequency corresponding to the $^2$H NMR timescale. Moreover, dynamical susceptibilities, which were obtained from Fourier-Laplace transformation of second rank orientational correlation functions from MD simulations and shifted along the frequency axis as described in the text, are included. The results for the end-to-end vector (MD end-to-end) and the C--H bonds of the phenyl ring (MD CH-ring) are shown as black dashed and dashed-dotted lines, respectively. The dotted curve represents the superposition of the scaled MD susceptibilities (see text). b) Mean relaxation times of the slow (solid blue dots) and the fast (solid green triangles) contribution to the DDLS relaxation peak together with mean rotational correlation times from $^{2}$H NMR (solid red squares) and MD simulations (open symbols).}
  \label{fig:figure3}
\end{figure}

Our quantitative DDLS analysis is exemplified based on the spectrum for $n = 8$ at 300\,K in Fig.\,\ref{fig:figure3} (a). The solid orange curve represents the model fit introduced in our previous work.\cite{zeissler2023influence}
In this model, the bimodal main relaxation peak is described by a superposition of two CD functions (blue and green solid curves). The vibrational contributions at higher frequencies, collectively referred to as microscopic dynamics, are modeled by a combination of a Brownian oscillator and a Debye function and are not shown as separate contributions in the figure. To properly take account of the suppression of the relaxation processes in the regime of microscopic dynamics, an inertial rise function is applied in the time domain to both CD components. This model has been shown to adequately describe DDLS spectra exhibiting bimodal relaxation behavior.\cite{zeissler2023influence}

Further insights are available when comparing the DDLS data with results from NMR experiments and MD simulations. In $^{2}$H NMR, we measure the SLR times $T_1$ for the phenyl-ring deuterons of the 1-phenylalkanes with $n=4$, $n=6$, and $n=8$ in broad temperature ranges, see \dag{SI}. The resulting correlation times are included in Fig.\ \ref{fig:figure3}. In panel (a), the vertical dashed line indicates the frequency corresponding to the mean correlation time directly obtained from $T_1$ for $n = 8$ at 300\,K using Eq.\ \eqref{eq:HT-Naeherung}. We see that the NMR time scale is very similar to that of the fast DDLS process. In MD simulations, we calculate second rank orientational correlation functions and obtain the corresponding dynamical susceptibilities after Fourier-Laplace transformation. In Fig.\ \ref{fig:figure3} (a), the dashed and dashed-dotted curves represent the susceptibilities resulting for the end-to-end vector and the C--H bonds of the phenyl ring, respectively. To eliminate mild differences in the absolute time scales of the computational and experimental data, both MD susceptibilities were shifted along the frequency axis by a constant factor $a=2.3$ such that the MD end-to-end vector susceptibility matches the peak position of the slow DDLS process. After this correction, the peak position of the MD phenyl-ring susceptibility is consistent with the NMR and fast DDLS time scales, indicating that the relative mobility of different molecular moieties is well reproduced in our MD approach.

Fig.\,\ref{fig:figure3} (b) shows the mean relaxation times $\tau_{\mathrm{mean,\,CD}}=\beta_{\mathrm{CD}}\tau_{\mathrm{CD}}$ of both DDLS processes, as obtained from the above described fit procedure, along with the mean rotational correlation times resulting from $^{2}$H NMR, see Eq.\ \eqref{eq:HT-Naeherung}, and peak correlation times from MD simulations. The CD width parameters in the fit of the DDLS data were fixed to the values obtained at 360\,K ($\beta_{\mathrm{CD,\,slow}}=0.71$ and $\beta_{\mathrm{CD,\,fast}}=0.79$ for the slow and the fast contributions, respectively) as it was done in ref.\,\citenum{zeissler2023influence}. In fact, the same fitting results as presented in ref.\,\citenum{zeissler2023influence} were used in the current work. As aforementioned, the MD correlation times are shifted by a constant factor $a^{-1}=0.44$ at all temperatures such that the peak time of the calculated end-to-end vector susceptibility agrees with that of the slow DDLS contribution at 300\,K. The time constants of the fast DDLS process agree with those from the $^{2}$H NMR analysis at all studied temperatures. Moreover, the timescale separation of the end-to-end vector and phenly-ring reorientation from MD simulations matches the separation of the fast and slow contributions to the DDLS relaxation peak. These results show that the fast relaxation process in DDLS can be assigned to the rotational dynamics of the phenyl ring. The situation is very similar for other alkyl chain lengths, with the timescale separation increasing with $n$. The case of $n=13$ is presented in the \dag{SI}.\\

\begin{figure}[th]
  \includegraphics[width=0.5\textwidth]{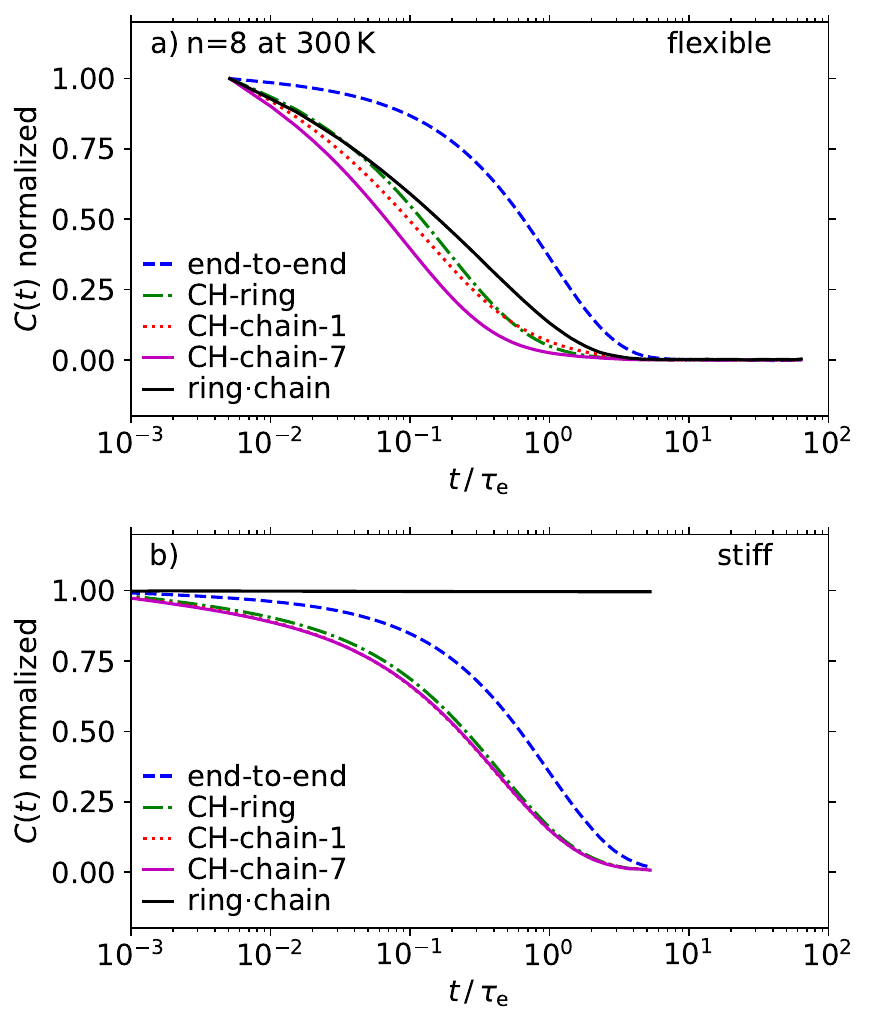}
  \caption{a) Second rank orientational correlation functions of different intramolecular vectors for $n=8$ at 300$\,$K from MD simulations. The data are shown on a reduced time scale $t/\tau_\text{e}$, where $\tau_\text{e}$ is the rotational correlation time of the end-to-end vector. The considered vectors are visualized in Fig.\,\ref{fig:figure1}. While the data for the phenyl ring (CH-ring) represents the average over all C--H
   bond vectors in this molecular moiety, we discriminate between C--H bond vectors at various positions along the alkyl tail. Specifically, we show results for bonds in the first (CH-chain-1) and the seventh (CH-chain-7) CH$_2$ unit of the tail, counting from the phenyl ring. b) The same second rank orientational correlation functions of the artificially stiffened molecule for $n=8$ at 300\,K.}
  \label{fig:figure4}
\end{figure}
Having demonstrated that our MD simulations capture the timescale separation of the two rotational modes, we use this approach to determine to what extent the faster reorientation of the phenyl ring results from internal rotation of this molecular entity and overall anisotropic rotation of the molecule, respectively. Fig.\,\ref{fig:figure4} (a) shows the calculated second rank orientational correlation functions of various intramolecular vectors for $n=8$ at 300\,K. In particular, we discriminate between C--H bonds in the phenyl ring (CH-ring) and at various positions along the alkyl tail, explicitly, between those in the first (CH-chain-1) and seventh (CH-chain-7) methylene groups, where we start counting at the phenyl ring. We see that the CH-ring and CH-chain-1 correlation functions decay on a very similar timescale, while C--H bonds close to the tail end show significantly faster reorientation. This result could potentially be interpreted in two ways. First, we may assume that an anisotropic rotation of an essentially rigid molecule results in a timescale separation of phenyl ring- and end-to-end vector rotation and that chain-end effects lead to enhanced dynamics in this part of the molecule. Second, we may suppose that the anisotropy of the overall reorientation is weak but the molecule has high internal flexibility and, in particular, the phenyl ring may reorient relative to the alkyl chain. 

To scrutinize these conjectures, we determine on which timescale C--H bonds in the first methylene group and in the nearby phenyl ring of the same molecule change their \emph{relative} orientation. Fig.\,\ref{fig:figure4} shows the corresponding normalized correlation function (ring$\cdot$chain) calculated based on Eq.\ \eqref{eq:introt}. While this correlation function will not decay if the molecule rotates as a rigid structure, we see that the relative orientation of the ring and tail bonds changes on a similar timescale as that of the individual bonds. For comparison, Fig.\,\ref{fig:figure4} (b) shows the same plot for the stiffened molecule, for which the intramolecular flexibility is suppressed by strong dihedral potentials (see Sec.\ \ref{subsec:MD}). In this case, the ring-chain correlation does not decay at all, the correlation functions of all C--H bonds collapse onto each other, and the timescale separation between bond-vector and end-to-end vector reorientation decreases but does not fully vanish. These results suggest that internal rotation of the phenyl ring is essential for the observed timescale separation. However, for the stiffened molecules, i.e., even without internal rotations, significant dynamic separation of end-to-end vector and ring rotation is observed, suggesting that similar phenomena as observed for the 1-phenylalkanes could also be observed in rigid anisotropically shaped molecules.

\begin{figure}[ht]
  \includegraphics[width=0.5\textwidth]{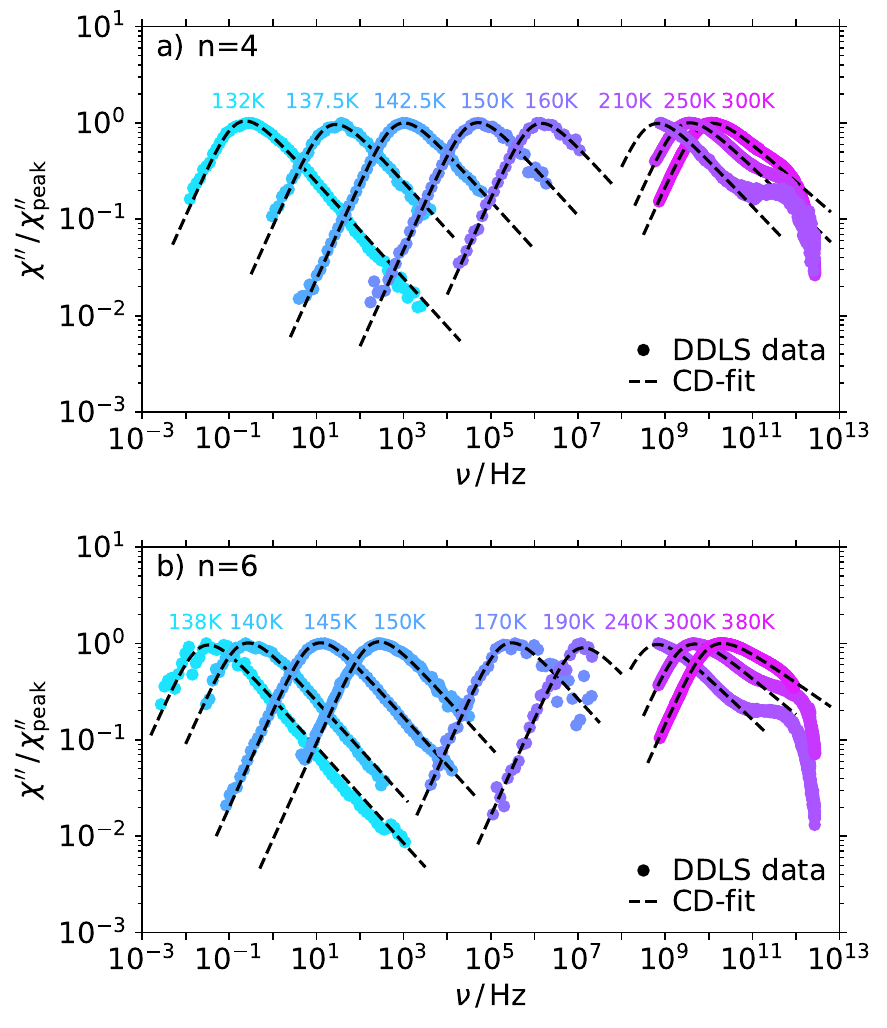}
  \caption{DDLS spectra for $n=4$ (a) and $n=6$ (b) from temperatures close to the glass transition up to temperatures far above the melting point (see text). Dashed curves represent fits to the peak regions by the Cole-Davidson model.}
  \label{fig:figure5}
\end{figure}

To investigate possible changes in the shape of the relaxation spectra upon supercooling, we next perform DDLS measurements in a much broader temperature range than in our previous studies.\cite{zeissler2023influence, zeissler2025relation} However, this approach is limited to compounds with $n < 8$ since 1-phenylalkanes with $n \geq 8$ tend to crystallize rapidly when cooled below their melting points. Fig.\,\ref{fig:figure5} shows DDLS spectra for $n = 4$ (a) and $n = 6$ (b), covering the full dynamic range, from temperatures near the glass transition temperature ($T_{\mathrm{g}} \sim 129\,\mathrm{K}$ for $n = 4$, and $T_{\mathrm{g}} \sim 135\,\mathrm{K}$ for $n = 6$) up to well above the respective melting points ($T_{\mathrm{m}} \sim 185\,\mathrm{K}$ for $n = 4$, and $T_{\mathrm{m}} \sim 212\,\mathrm{K}$ for $n = 6$).
In the deeply supercooled regime, the spectra exhibit a well-resolved relaxation peak that shifts to higher frequencies with increasing temperatures until it eventually overlaps with vibrational contributions. The corresponding relaxation times were obtained by fitting the peak region using the Cole–Davidson model (dashed curves in Fig.\,\ref{fig:figure5}) with fixed width parameters $\beta_{\mathrm{CD}}=0.5$ . We note that spectra at lower temperatures, explicitly for peak frequencies below $10^7\,\mathrm{Hz}$, were measured by PCS, while those at higher temperatures were obtained using TFPI. Due to a gap between the accessible frequency ranges of these two techniques ($\sim$10$^7$–$10^8\,\mathrm{Hz}$), there is a temperature range, typically around the melting point, where the relaxation peak cannot be fully resolved.

\begin{figure}[ht!]
  \includegraphics[width=0.5\textwidth]{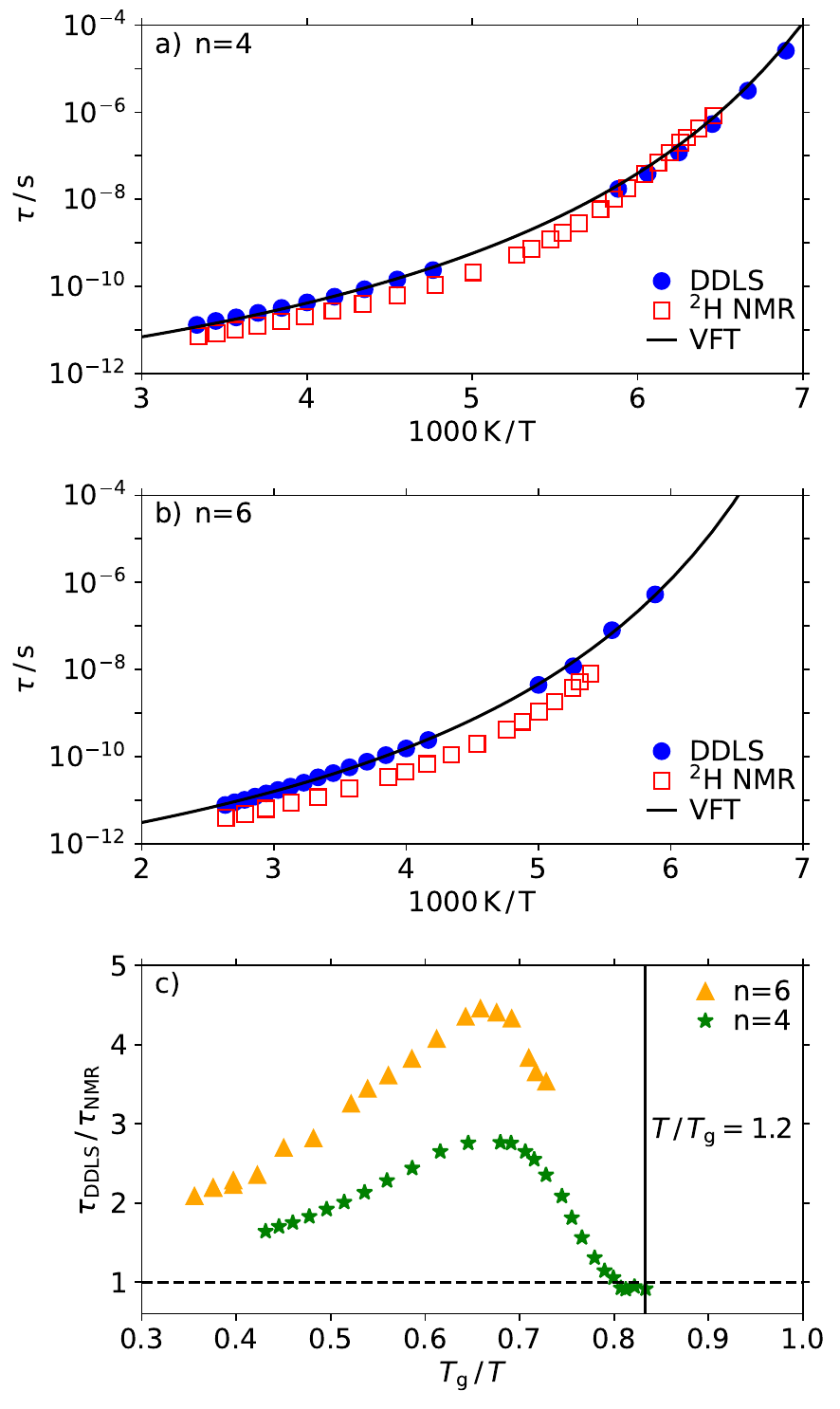}
  \caption{Peak correlation times of $n=4$ (a) and $n=6$ (b) from DDLS and $^{2}$H NMR of the ring-labeled liquids. The black curve is a VFT fit of the DDLS relaxation times. c) Separation of DDLS and NMR timescales, i.e., $\tau_{\mathrm{DDLS}}/\tau_{\mathrm{NMR}}$ in dependence on the rescaled inverse temperature $T_{\mathrm{g}}/T$. The horizontal black dashed line marks the merging of timescales at $T/T_{\mathrm{g}}\sim1.2$.}
  \label{fig:figure6}
\end{figure}

Consistent with the observations in Fig.\,\ref{fig:figure2}, we see in Fig.\,\ref{fig:figure5} that 1-phenylalkanes with $n \leq 6$ do not exhibit a clearly visible bimodality in their DDLS relaxation peaks. However, at sufficiently high temperatures, a slight shoulder on the high-frequency side of the peak suggests the presence of a second process. To study the effect in more detail, we again compare DDLS and $^{2}$H NMR results. As was outlined in Sec.\ \ref{sec:NMR}, NMR correlation times in a broader temperature range are available from $^2$H $T_1$ data when using a Cole-Davidson spectral density in Eq.\ \eqref{eq:bpp}. Specifically, the Cole-Davidson parameters $\tau_\text{CD}$ and $\beta_\text{CD}$ from such a $^2$H SLR analysis can be used to obtain peak correlation times for straightforward comparison with DDLS results.

Figures \ref{fig:figure6} (a) and (b) show the peak correlation times obtained from DDLS and $^{2}$H NMR for $n = 4$ and $n = 6$, respectively. The black curves represent fits to the DDLS data using the Vogel–Fulcher–Tammann (VFT) equation. We see that the $^{2}$H NMR timescale is shorter than the DDLS timescale at $T>T_\text{m}$ and in the weakly supercooled liquid regime, while both timescales approach each other upon further cooling and merge at $\tau \sim 10^{-7}$\,s.

Fig.\,\ref{fig:figure6} (c) displays the ratio of the DDLS and $^{2}$H NMR timescales, $\tau_{\mathrm{DDLS}}/\tau_{\mathrm{NMR}}$, as a function of $T_{\mathrm{g}}/T$. The DDLS timescale needed to be taken from the VFT fits presented in Figs.\,\ref{fig:figure6} (a) and (b) since there is no exact match of the temperatures studied in DDLS and $^{2}H$ NMR. For both $n = 4$ and $n = 6$, the timescale separation initially increases upon cooling, reaches a maximum, and then decreases until the timescales eventually merge at $T/T_{\mathrm{g}} \sim 1.2$ in the case of $n = 4$. For $n = 6$, the latter dynamic regime could not be fully accessed because crystallization interfered in the NMR measurements; however, the available data suggest a similar trend. These findings indicate that the separation between the two rotational timescales diminishes upon supercooling. It should be noted that $T/T_{\mathrm{g}}=1.2$ is not an unexpected temperature for such a transition to occur. Typically, in this temperature range the temperature traces of $\alpha$- and $\beta$-relaxation times split for many supercooled liquids \cite{rossler1990indications} and a decoupling of translational and rotational
\begin{figure}[h]
  \includegraphics[width=0.5\textwidth]{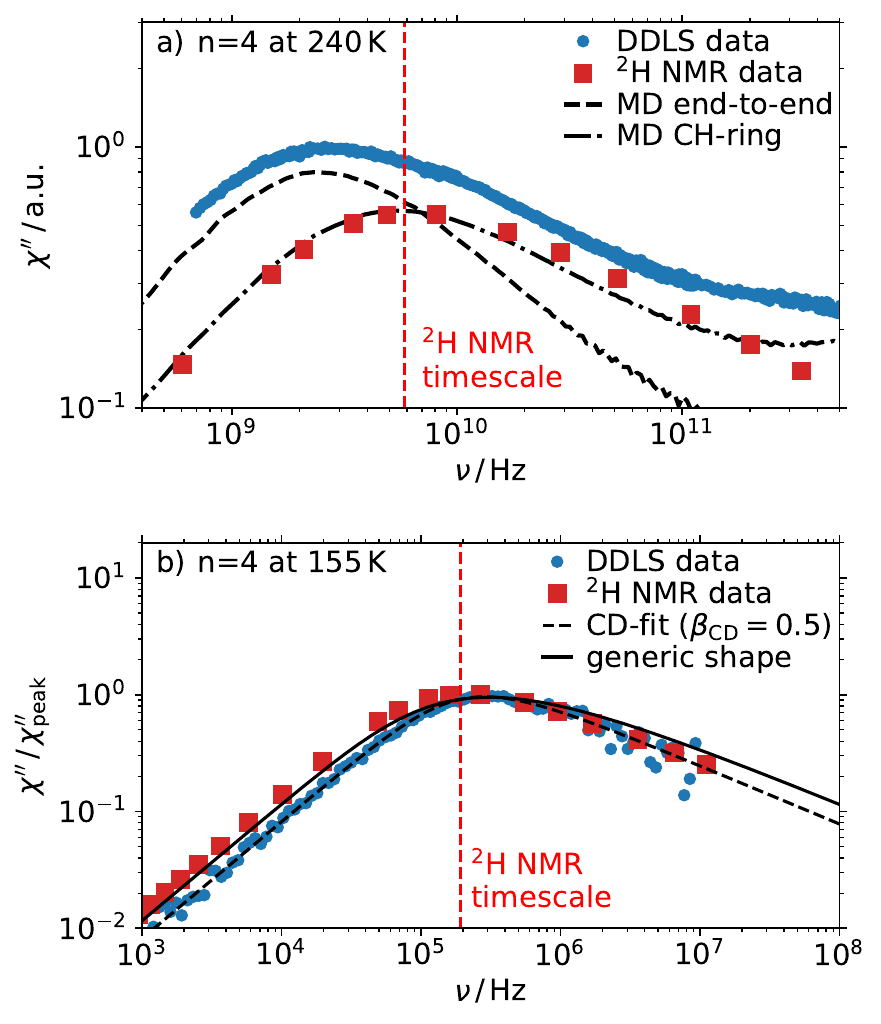}
  \caption{a) DDLS spectrum for $n=4$ at 240\,K. The dashed and dashed-dotted curves represent the susceptibilities obtained from MD simulations and shifted as described in the text. The vertical dashed line corresponds to the $^{2}$H NMR correlation time at this temperature and the red squares represent the corresponding NMR susceptibility (see text). b) DDLS spectrum for $n=4$ at 155\,K. The dashed black curve denotes a fit of the DDLS data with the Cole-Davidson model and the solid black curve represents the generic shape identified in relaxation spectra of a number of molecular glass formers close to the glass transition. NMR timescale and susceptibility are presented analogous to part (a) of the figure.}
  \label{fig:figure7}
\end{figure}
dynamics sets in \cite{rossler1990indications, chang1997heterogeneity, swallen2003self}.

To further elucidate the temperature-dependent evolution of the susceptibility shape, we compare DDLS, NMR, and MD results at characteristic temperatures. Fig.\,\ref{fig:figure7} (a) shows the DDLS spectrum for $n = 4$ at 240\,K, i.e., at $T\gg1.2\,T_\text{g}$, together with the MD susceptibilities again shifted by a constant factor $b=2.6$ and shifted in amplitude for visualization purposes. The vertical dashed line corresponds to the $^{2}$H NMR correlation time at this temperature and the red squares the corresponding NMR susceptibility $\chi^{\prime\prime}_{\mathrm{NMR}}$.
Exploiting the relation $\chi^{\prime\prime}(\omega) = \omega J(\omega)$, the NMR susceptibility can be obtained from Eq.~\eqref{eq:bpp} as 
\begin{equation}
    \chi^{\prime\prime}_{\mathrm{NMR}}(\nu)=\frac{2\pi\nu}{T_1}\ .
\end{equation}
$T_1$ measurements were performed at a single Larmor frequency and different temperatures.
To get the frequency dependence of $\chi^{\prime\prime}_{\mathrm{NMR}}(\nu)$ we utilize the known relation between $\tau_\text{NMR}$ and temperature, shown in Fig. \ref{fig:figure6} a) under the assumption that frequency-temperature superposition (FTS) holds in the considered temperature range. By multiplying $\tau_\text{NMR}(T)$ with $\omega_\text{L}$ we get a normalized x-axis with the susceptibility peak at 0.6. The such normalized values can subsequently be scaled on the peak frequency of a specific temperature to enable comparisons between DDLS, MD simulations and NMR for this temperature.
This comparison shows that the NMR and MD susceptibilities of the phenyl-ring bonds have a very similar shape: Moreover, it confirms that DDLS, NMR, and MD results for the time scale separation of the fast and slow processes agree. Fig.\,\ref{fig:figure7} (b) shows the situation at 155\,K, i.e., at $T\sim1.2\,T_\text{g}$, where the DDLS and $^{2}$H NMR time constants have already fully merged. Consistently, the NMR and DDLS susceptibilities peaks occur at very similar frequencies. At this lower temperature, the DDLS relaxation peak does not show any sign of a high frequency shoulder anymore and is well described by the Cole-Davidson model with width parameter $\beta_{\mathrm{CD}}=0.5$. Interestingly, the NMR susceptibility is close to identical in peak shape. The solid black curve represents the generic shape identified in relaxation spectra of a number of molecular glass formers close to the glass transition temperature. \cite{pabst2021generic,bohmer2025spectral} Our analysis shows that the influence of anisotropic rotation and internal degrees of freedom diminishes upon supercooling in the case of liquid 1-phenylalkanes, resulting in a merging of the timescales of molecular rotation observed by DDLS and NMR as well as a convergence of susceptibility peak shapes.

\section{Conclusions}

By combining results from DDLS, $^{2}$H NMR, and MD simulations, we have demonstrated that the observed bimodality of the DDLS relaxation peaks in 1-phenylalkanes mainly originates from internal rotation of the phenyl ring. In particular, the timescale separation of the two contributions to the DDLS relaxation peaks is very similar to that of ring rotation and end-to-end vector rotation in the MD simulations. Moreover, $^{2}$H NMR experiments on ring-deuterated molecules yielded correlation times, which agreed with those of the fast DDLS process over a wide temperature range, confirming that phenyl-ring reorientation is at the origin of the latter rotational mode.

This insight enabled the use of NMR to detect the bimodal character of the DDLS relaxation peaks even in short-chained 1-phenylalkanes, where the bimodality is not readily apparent from DDLS measurements alone, allowing us to investigate this effect deep into the supercooled regime. Comparison of DDLS and $^{2}$H NMR correlation times revealed that differences between the reorientation dynamics of various moieties of flexible molecules diminish upon cooling. To rationalize this effect, we propose that the high cooperativity of molecular dynamics in deeply supercooled liquids makes individual molecular characteristics disappear and leads to common reorientation dynamics of larger entities.

Since 1-phenylalkanes serve as simple model liquids with molecular features common to many other molecular liquids, we suggest that the diverse relaxation behaviors observed across different molecular liquids at $T\gtrsim T_\text{m}$ are largely caused by similar mechanisms. Conversely, the generic relaxation behavior seen in the supercooled regime may result from the suppression of these effects due to increased cooperativity of the dynamics at lower temperatures.

\section*{Conflicts of interest}

There are no conflicts to declare.

\section*{Acknowledgements}
Financial support by the Deutsche Forschungsgemeinschaft under grant no.\ 1192/3 is gratefully acknowledged.

\renewcommand\refname{References}

\bibliography{ringtail.bib}
\end{document}